\documentclass{ws-ijmpa}
\usepackage{bm}% bold math
\newcommand{\beq}{\begin{equation}}
\newcommand{\eeq}{\end{equation}}
\newcommand{\beqa}{\begin{eqnarray}}
\newcommand{\eeqa}{\end{eqnarray}}
\def\nn{\nonumber\\}
\def\eq#1{(\ref{#1})}

\def\cd#1{\ensuremath{\nabla_{#1}}}          %covariant derivative
\def\pd#1{\ensuremath{\partial_{#1}}}        %partial derivative
\def\st{space-time}

\def\lab#1{\label{#1}}

\def\altura{\rule{0pt}{16pt}}
\begin{document}

\markboth{Ricardo E.  Gamboa Sarav\'{\i}}
{The Gravitational Field of a Plane Slab}

\catchline{}{}{}{}{}
%
%%%%%%%%%%%%%%%%%%%%%%%%%%%%%%%%%%%%%%%%%%%%%%%%%%%%%%%%%%%%%%%%%%%%

\title{The Gravitational Field of a Plane Slab
}

\author{RICARDO E. GAMBOA SARAV\'I}

\address{Departamento de
F\'\i sica, Facultad de Ciencias Exactas,\\ Universidad Nacional de
La Plata and IFLP,
CONICET.\\C.C. 67, 1900 La Plata, Argentina,
\\
quique@fisica.unlp.edu.ar}

\maketitle

\begin{history}
\received{Day Month Year}
\revised{Day Month Year}
\end{history}

%%%%%%%%%%%%%%%%%%%%%%%%%%%%%%%%%%%%%%%%%%%%%%%%%%%%%%%%%%%%%%%%%%%%%%%
\begin{abstract}
We discuss  the exact  solution of Einstein's equation
corresponding to a static and plane symmetric distribution of
matter with constant positive density located below $z=0$ matched
to vacuum solutions. The internal solution depends essentially on
two constants: the density $\rho$ and a parameter $\kappa$. We
show that these space-times finish down below at an inner
singularity at finite depth $d\leq\sqrt{\frac{\pi}{24\rho}}$.  We
show that for $\kappa\geq0.3513\dots$, the dominant energy
condition is satisfied all over the space-time.

We match these singular solutions to the vacuum one and compute
the external gravitational field in terms of  slab's parameters.
Depending on the value of $\kappa$, these slabs are either
attractive, repulsive or neutral. The external solution turns out
to be  a Rindler's \st. Repulsive slabs explicitly show how
negative, but finite pressure can dominate the attraction of the
matter. In this case, the presence of horizons in the vacuum shows
that there are null geodesics which never reach the surface of the
slab.

We also consider  a static and plane  symmetric non-singular
distribution of matter with constant positive density $\rho$ and
thickness $d$ ($0<d<\sqrt{\frac{\pi}{24\rho}}$) surrounded by  two
external vacuums. We explicitly write down the pressure and the
external gravitational fields in terms of $\rho$ and  $d$. The
solution turns out to be attractive and remarkably  asymmetric:
the ``upper" solution is Rindler's vacuum, whereas the ``lower"
one is the singular part of Taub's plane symmetric solution.
Inside the slab, the pressure is positive and bounded, presenting
a maximum at an asymmetrical position between the boundaries. We
show that if $0<\sqrt{6\pi\rho}\,d<1.527\dots$, the dominant
energy condition is satisfied all over the space-time. We also
show how the mirror symmetry is restored at the Newtonian limit.

We also find thinner repulsive slabs by matching a singular slice
of the inner solution to the vacuum.

We also discuss  solutions in which  an attractive slab and a
repulsive one, and two neutral ones are joined. We also discuss
how to assemble a ``gravitational capacitor" by inserting a slice
of vacuum between two such slabs.
\end{abstract}

%%%%%%%%%%%%%%%%%%%%%%%%%%%%%%%%%%%%%%%%%%%%%%%%%%%%%%%%%%%%%%%%%%%%%%%%%%%%%%%%%%%%%%%%%%%%

\section{Introduction}

Due to  the complexity of Einstein's field equations, one cannot
find exact solutions except in spaces of rather high symmetry, but
very often with no direct physical application. Nevertheless,
exact solutions can give an idea of the qualitative features that
could arise in General Relativity,  and so, of possible properties
of realistic solutions of the field equations.

 We have recently discussed exact solutions of Einstein's equation
presenting an {\em empty} (free of matter)  singular repelling
boundary  \cite{gs,gs1}. These singularities are not the sources
of the fields, but they arise owing to the attraction of distant
matter.

In this paper, we want to illustrate this and other  curious
features of relativistic gravitation by means of a simple exact
solution:  the gravitational field of a static plane symmetric
relativistic perfect incompressible  fluid with positive density
located below $z=0$ matched to  vacuum solutions.  In reference
\cite{gs2}, we analyze in detail the properties of this internal
solution, originally  found by A. H. Taub \cite{taub2} (see also
\cite{AH,NKH,SKMHH}), and we find that it finishes up down below
at an inner singularity at finite depth $d$, where
$0<d<\sqrt{\frac{\pi}{24\rho}}$. Depending on the value of a
parameter $\kappa$, it turns out to be gravitational attractive
($\kappa<\kappa_{\text{crit}}$), neutral
($\kappa=\kappa_{\text{crit}}$) or repulsive
($\kappa>\kappa_{\text{crit}}$), where
$\kappa_{\text{crit}}=1.2143\dots$. We also show that for
$\kappa\geq0.3513\dots$, the dominant energy condition is
satisfied all over the space-time.

In this paper, we make  a detailed analysis  of the matching of
these exact solutions to vacuum ones. Here, we impose the
continuity of the metric components and of their first derivatives
at the matching surfaces,  in contrast to reference \cite{gs2},
where not all these derivatives are continuous at the boundary.

In the first place, we consider the matching of the whole singular
slabs to the vacuum, and explicitly compute  the external
gravitational fields in terms of the slab parameters. Repulsive
slabs explicitly show how negative but finite pressure can
dominate the attraction of the matter. In this case, they have the
maximum depth, i.e., $d=\sqrt{\frac{\pi}{24\rho}}$, and the
exterior solution presents horizons showing that there are
vertical photons that cannot reach the slab surface.

Secondly, we  consider  a  non-singular slice of these slabs with
thickness $d$ ($0<d<\sqrt{\frac{\pi}{24\rho}}$) surrounded by two
external vacuum. Some of the properties of this solution have
already been discussed in reference \cite{NKH}. Here, we
explicitly write down the pressure and the external gravitational
fields in terms of $\rho$ and  $d$. The solution turns out to be
attractive, and remarkably asymmetric: the ``upper" solution is
Rindler's vacuum, whereas the ``lower" one is the singular part of
Taub's plane symmetric solution. Inside the slab, the pressure is
positive and bounded, presenting a maximum at an asymmetrical
position between the boundaries. We show that if
$0<\sqrt{6\pi\rho}\,d<1.527\dots$, the dominant energy condition
is satisfied all over the space-time. This solution finishes up
down below at an empty repelling boundary  where space-time
curvature diverges. This exact solution clearly shows how the
attraction of distant matter can shrink the \st\ in such a way
that it finishes at a  free of matter singular boundary, as
pointed out in \cite{gs}. We also show how the mirror symmetry is
restored at the Newtonian limit.

We also construct thinner repulsive slabs by matching a singular
slice of the inner solution to vacuum. These slabs turn out to be
less repulsive than the ones discussed above, since all incoming
vertical null geodesics reach the slab surface in this case.

For the sake of completeness, in section \ref{solution},  we
include  some  results from reference \cite{gs2} which are
necessary for the computations of the following sections. In
section \ref{dec}, we show under which conditions  the dominant
energy condition is satisfied. In section \ref{match}, we discuss
how solutions can be matched. In section \ref{whole}, we study the
matching of the whole singular interior solution to vacuum. In
section \ref{dos}, we match  two interior solutions facing each
other. In  section \ref{doble}, we discuss the matching of a
non-singular slice of the interior solution with two different
vacuums. In section \ref{asymmetry}, we show how the mirror
symmetry of this solution is restored at the Newtonian limit. In
section \ref{repulsive} we construct thinner repulsive slabs
($d<\sqrt{\frac{\pi}{24\rho}}$) by matching a singular slice of
the inner solution to vacuum.

Throughout this paper, we adopt the convention in which the \st\
metric has signature $(-\ +\ +\ +)$, the  system of units in which
the speed of light $c=1$,  Newton's gravitational constant $ G=1$
and $g$  denotes gravitational field and not the determinant of
the metric.

\section{The interior solution} \lab{solution}

 In this section, we consider  the solution  of Einstein's
equation corresponding to  a static and plane symmetric
distribution of matter with constant positive density and plane
symmetry. That is, it must be invariant under translations in the
plane and under rotations around its normal. The matter we shall
consider is a perfect fluid of uniform density $ \rho$. The
stress-energy tensor is \beq T_{ab}= (\rho+p)\,u_au_b+p\, g_{ab}\,
,\eeq where $u^a$ is the velocity of fluid elements.

Due to the plane symmetry and staticity, following \cite{taub} we
can find coordinates $(t, x, y, z)$ such that \beq \lab {met}
ds^2= - \mathcal{G}(z)^{2 }\ dt^2+ e^{2V(z)}\left(dx^2+dy^2
\right)+dz^2\,,\eeq%
that is,  the more general metric admitting the Killing
vectors $\pd x$, $\pd y$, $x\pd y-y\pd x$ and $\pd t$.

The non identically vanishing components of the Einstein tensor
are
\beqa \lab {gtt} G_{tt}=-\,\mathcal{G}^2 \left(2\,V''+3\, V'^2\right)\\
\lab {gii} G_{xx}=G_{yy}= e^{2V} \left({\mathcal{G}''}/{\mathcal{G}}+{\mathcal{G}'}/{\mathcal{G}}\;V'+V''+V'^2\right)\,,\\
\lab {gzz} G_{zz}=  V' \left(2\ {\mathcal{G}'}/{\mathcal{G}}+  V'
\right) ,
 \eeqa where a prime $(')$ denotes differentiation with respect to $z$.%

On the other hand,   due to the assumed symmetries and to the fact
that the material content is a perfect fluid,
$u_a=(-\mathcal{G},0,0,0)$,  so \beq T_{ab}=
\text{diag}\left(\rho\, \mathcal{G}^{2},p\, e^{2V},p\,
e^{2V},p\right)\,, \eeq where
 $p$  depends only on the z-coordinante.
 Thus, Einstein's equations, i.e., $G_{ab}=8 \pi T_{ab}$, are
\beqa \lab {gtt1} 2\,V''+3\, V'^2= -8 \pi{\rho}\,, \\%
\lab
{gii1}{\mathcal{G}''}/{\mathcal{G}}+{\mathcal{G}'}/{\mathcal{G}}\;V'+V''+V'^2
=
8 \pi{p}\, ,\\
\lab {gzz1}  V' \left(2\ \mathcal{G}'/{\mathcal{G}}+  V' \right) =
8 \pi{p}\,.
 \eeqa%

Moreover, $\cd a T^{ab}=0$ yields
\beq \lab{ppr} p' = -(\rho+p)\,\mathcal{G}'/\mathcal{G}\,.\eeq Of
course, due to Bianchi's identities equations, (\ref{gtt1}),
(\ref{gii1}), (\ref{gzz1}) and (\ref{ppr}) are not independent, so
we shall here use  only \eq{gtt1}, \eq{gzz1}, and \eq{ppr}.

Since $\rho$ is constant, from \eq {ppr} we readily find
 \beq\lab{p}p  = C_p /\mathcal{G}(z)-\rho,  \eeq where $C_p$ is an arbitrary constant.

By setting %\beq \lab{W}
$W(z)= e^{\frac{3}{2}V(z)}$, %\eeq
we can write  \eq{gtt1} as $W''=-{{6\pi\rho}}\, W$, and its
general solution  can be written as \beqa \lab{W1}W(z)= C_1\,
\sin{(\sqrt{{6\pi\rho}}\ z+C_2)},\, \eeqa
% \beqa \lab{W1} W(z)= \cases{C_1\, \frac{\sin\sqrt{\alpha}(z+C_2)}{\sqrt{\alpha}}&{if}\,\,\,\ %$\rho>0$\\
% C_1\, z+C_1\,C_2&{if}\,\,\,\ $\rho=0$\\C_1\, %\frac{\sinh\sqrt{-\alpha}(z+C_2)}{\sqrt{-\alpha}}&{if}\,\,\,\ $\rho<0$\ ,}\eeqa %
where $C_1$ and $C_2$ are arbitrary constants. %Notice that the
%three cases in \eq{W1} are covered (as a limit when $\alpha=0$)
%by the first one.
Therefore,  we have
 \beq \lab{V} V(z)={\frac{2}{3}}\,\ln\left(
{C_1\,\sin{(\sqrt{{6\pi\rho}}\ z+C_2)}}\right).\eeq

 Now, by replacing \eq {p} into \eq {gzz1}, we get the  first order linear differential equation which $\mathcal{G}(z)$ obeys
\beqa  \lab {gzz2} \mathcal{
G}'=-\left(\frac{4\pi\rho}{V'}+\frac{V'}{2}\right)\mathcal{G} +
\frac{4\pi C_p}{V'}\ \\ =-{\sqrt{{6\pi\rho}}}\left(\tan
u+\frac{1}{3}\cot u\right)\mathcal{G} +{\sqrt{{6\pi\rho}}}\,\
\frac{ C_p}{\rho}\tan u \,,\eeqa where $u=\sqrt{{6\pi\rho}}\
z+C_2$. And in the last step, we have made use of \eq {V}. The
general solution of \eq {gzz2} can be written as \footnote {In the
appendix we show  how  the integral appearing in the first line of
\eq{integral} is performed. }
\beqa\lab{integral}  \mathcal{G}=  \frac{\cos u}{(\sin u)^{1/3}}\left(C_3 \ +\frac{C_p}{\rho}\int_0^u\frac{(\sin u')^{\frac{4}{3}}}{(\cos u')^2}du'\right)\nn \lab {G} = C_3\, \frac{\cos u}{\left(\sin u \right)^{\frac{1}{3}}}+\frac{3 C_p}{7\rho}\  {\sin^2\! u}\,\, _2F_1\!\Bigl(1,\frac{2}{3};\frac{13}{6};\sin^2u \Bigr), \eeqa %
where $C_3$ is another arbitrary constant, and $_2F_1(a,b;c;z)$ is
the Gauss hypergeometric function (see the appendix at the end of
the paper).

Therefore,  the line element \eq{met} becomes%
 \beqa \lab{met1}
ds^2=  -  \mathcal{G}(z)^{2}\,
 dt^2 + \left({C_1\,\sin u}\right)^{\frac{4}{3}}\left(dx^2+dy^2\right) + dz^2,
 \eeqa%
 where $\mathcal{G}(z)$ is given in \eq {G} and  $u=\sqrt{{6\pi\rho}}\ z+C_2$.
Thus, the solution contains five arbitrary constants: $\rho$,
$C_p$, $C_1$, $C_2$, and $C_3$. The range of the coordinate $z$
depends on the value of these constants.

Notice  that the metric \eq{met1} has a \st\, curvature
singularity where $\sin u=0$, since straightforward computation of
the scalar quadratic in the
Riemann tensor yields%
\beqa\lab{RR} R_{abcd}R^{abcd}=4
\left({\mathcal{G}''^2}+2\,{\mathcal{G}'^2}\,V'^2\right)/{\mathcal{G}^2}+4\left(2\,V''^2+4\,V''V'^2+3\,V'^4\right)\nn
  =\frac{256}{3}\,\,\pi^2\rho^2\,\left(2+{\sin^{-4} u}+ \frac{3}{4} \left(\frac{p}{\rho}+1\right)\left(\frac{3p}{\rho}-1\right)\right), \eeqa
so $ R_{abcd}R^{abcd}\rightarrow\infty$ when $\sin u\rightarrow0$.

On the other hand, by contracting Einstein's equation, we get \beq
\lab{R} R(z)=8\pi(\rho-3p(z))=8\pi(4\rho-3C_p
/\mathcal{G}(z))\,.\eeq

For  $\rho>0$, $C_p>0$  and $C_3>0$, the solution \eq{met1} was
found by Taub  \cite{taub2,SKMHH}. Nevertheless, this solution
has a  wider range of validity.

Of course, from this solution we can obtain  vacuum ones as a
limit. In fact, when $C_p =0$, it is clear from \eq{p} that
$p(z)=-\rho$, and the solution \eq{met1} turns out to be a vacuum
solution with a cosmological constant $\Lambda=8\pi\rho$
\cite{NH,SKMHH} \beqa \lab{metNH} ds^2=  -
{\cos^2u}\;\sin^{-\frac{2}{3}} u  \, dt^2 + \,\sin^{\frac{4}{3}} u
\,\left(dx^2+dy^2\right) + dz^2,\nn
-\infty<t<\infty,\quad-\infty<x<\infty,\quad-\infty<y<\infty,\quad0<u<\pi, \eeqa%
where $u=\sqrt{3\Lambda}/2\ z+C_2$.  We get from \eq{R} that it is
a space-time with constant scalar curvature $4\Lambda$, and from
\eq{RR} we get that \beqa\lab{RR2}
R_{abcd}R^{abcd}=\frac{4}{3}\,\,\Lambda^2\,\left(2+\frac{1}{\sin^4
u}\right)\ . \eeqa

Now, we  take the limit $\Lambda\rightarrow0$
($\rho\rightarrow0$). By setting $C_2=\pi-\frac{\sqrt{3\Lambda}}{6
g}$ and an appropriate rescaling of the coordinates
$\left\{t,x,y\right\}$, we can readily see that, when
$\Lambda\rightarrow0$, \eq {metNH} becomes \beqa \lab{taub} ds^2=-
(1-3gz)^{-\frac{2}{3}}\,
 dt^2 + (1-3gz)^{\frac{4}{3}}\left(dx^2+dy^2\right) +
 dz^2,\nn
-\infty<t<\infty,\quad-\infty<x<\infty,\quad-\infty<y<\infty,\quad0<1-3gz<\infty \,,  \eeqa%
where $g$ is an arbitrary constant.  In \eq{taub}, the coordinates
have been chosen in such a way that it describes a homogeneous
gravitational field $g$ pointing in the negative $z$-direction in
a neighborhood of  $z=0$. The metric \eq{taub} is Taubs's
\cite{taub} vacuum plane solution expressed  in the coordinates
used in Ref. \cite{gs}, where a detailed study of it can be found.

On the other hand, by setting
$C_2=\frac{\pi}{2}+\frac{\sqrt{3\Lambda}}{2 g}$ and an appropriate
rescaling of the coordinate $t$, we can readily see that, when
$\Lambda\rightarrow0$, \eq {metNH} becomes \beqa \lab{rindler}
ds^2=- (1+gz)^{{2}}\,
 dt^2 + dx^2+dy^2 +
 dz^2,\nn
-\infty<t<\infty,\quad-\infty<x<\infty,\quad-\infty<y<\infty,\quad-\frac{1}{g}<z<\infty \,,  \eeqa%
where $g$ is an arbitrary constant,  and the coordinates have been
chosen in such a way that it also describes a homogeneous
gravitational field $g$ pointing in the negative $z$-direction in
a neighborhood of  $z=0$. The metric \eq{rindler} is, of course,
Rindler's flat \st.

%Nevertheless, the solution \eq{met1} has a  wider range of validity.
 For {\em exotic  } matter, some interesting
solutions also arise, but  the complete analysis turns out to be
somehow involved. So, for the sake of clarity, we shall confine
our attention to positive values of $\rho$ and $C_p \neq0$,
leaving the complete study to a forthcoming publication
\cite{gs3}.

%\section{The properties of the function $\mathcal{G}(z)$}\label{properties}

%In this section we shall study in detail the properties of the
%solution for the case $\rho>0$ and $C_p \neq0$.

Now, it  is clear from (\ref{gtt1}), (\ref{gii1}), (\ref{gzz1})
and (\ref{ppr})
 that field equations are invariant under the transformation
  $z\rightarrow \pm z+z_0$, i.e., z-translations and mirror reflections
across any plane $z\!=$const. Thus, if
$\{\mathcal{G}(z),V(z),p(z)\}$ is a solution  $\{\mathcal{G}(\pm
z+z_0),V(\pm z+z_0),p(\pm z+z_0)\}$ is another one, where $z_0$ is
an arbitrary constant. Therefore, taking into account that
${u=\sqrt{{6\pi\rho}}\,z+C_2}$, without loss of generality, the
consideration of the case  $0<u<\pi/2$  shall suffice.

By an appropriate rescaling of the coordinates
$\left\{x,y\right\}$, without loss of generality,   we can write
the metric  \eq {met1} as \beqa \lab{met2} ds^2=  -
\mathcal{G}(z)^{2}\, dt^2 + \,\sin^{\frac{4}{3}}
u\,\left(dx^2+dy^2\right) + dz^2,\nn
-\infty<t<\infty,\quad-\infty<x<\infty,\quad-\infty<y<\infty,\quad0<u=\sqrt{{6\pi\rho}}\ z+C_2\leq\pi/2, \nn\eeqa%
and \eq{G} as \beqa \lab{G8} \mathcal{G}(z)=  \frac{\kappa C_p}{ \rho}\, \frac{\cos u}{\sin^{\frac{1}{3}} u }+\frac{3   C_p}{7 \rho}\  {\sin^{2} u}\,\,\, _2F_1\!\Bigl(1,\frac{2}{3};\frac{13}{6};\sin^2u \Bigr), \eeqa %
where $\kappa$ is an arbitrary constant.

 By replacing \eq{G8} into \eq {p}, we see that
%\beqa \lab {p2} p(u)= \rho \left(\frac{ \sin^{1/3} u }{{\kappa }\, \cos u +\frac{3}{7}\ %{\sin^{7/3}\!u}\,\,\, _2F_1\!\left(1,\frac{2}{3};\frac{13}{6};\sin^2u \right)}-1\right), \eeqa
the pressure is independent of $C_p$. On the other hand, since
$\mathcal{G}(z)$ appears squared in \eq{met2}, it suffices to
consider $C_p>0$. Therefore, rescaling  the coordinate $t$, we may
set $C_p=\rho$. Thus, \eq {G8} becomes \beqa \lab{G2}
\mathcal{G}(z) =G_\kappa(u)= \kappa \, \frac{\cos u}{\sin^{1/3} u
}+
\frac{3  }{7 }\  {\sin^{2} u}\,\,\, _2F_1\!\Bigl(1,\frac{2}{3};\frac{13}{6};\sin^2u \Bigr), \eeqa %
where $G_\kappa(u)$ is defined for future use, and we recall that
${u=\sqrt{{6\pi\rho}}\,z+C_2}$.  Furthermore, \eq {p} becomes
\beq\lab{p3}p(z)  = \rho \left(1 /\mathcal{G}(z)-1\right). \eeq

Therefore, the solution depends on two essential parameters,
$\rho$ and $\kappa$. We shall discuss in detail the properties of
the functions $\mathcal{G}(z)$ and $p(z)$ depending on the value
of the constant $\kappa$.

By using the transformation \eq{a1}, we can write $\mathcal{G}(z)$
as \beqa \lab{GH} \mathcal{G}(z) =
\frac{\left(\kappa-\kappa_{\text{crit}}\right){\cos u}+ \,
_2F_1\!\Bigl(-\frac{1}{2},-\frac{1}{6};\frac{1}{2};\cos^2u
\Bigr)}{{\sin^{1/3} u }}\,, \eeqa where \beq \lab{kapa}
\kappa_{\text{crit}}=\frac{\sqrt{\pi}\,{\Gamma(7/6)}}{{\Gamma(2/3)}}=1.2143\dots\,,\eeq
which is the form used in references \cite{AH,NKH}, and which is
more suitable to analyze its properties near $u=\pi/2$.

Now, the hypergeometric function in \eq{G2}   is a monotonically
increasing continuous positive function of $u$ for $0\leq u\leq
\pi/2$, since $c-a-b={1}/{2}>0$. Furthermore, taking  into account
that $_2F_1(a,b;c;0)=1$ and \eq{a1}, we have \beqa
_2F_1\!\Bigl(1,\frac{2}{3};\frac{13}{6};0\Bigr)=1,\,\,\,\,\text{and}\,
\,\,\,_2F_1\!\Bigl(1,\frac{2}{3};\frac{13}{6};1\Bigr)=
\frac{7}{3}. \eeqa Therefore, we readily see from \eq {G2} that,
no matter what the value of $\kappa$ is,
$\mathcal{G}(z)|_{u=\pi/2}=1$, and we get then from \eq {p3} that
$p(z)$  vanishes at $u=\pi/2$. On the other hand, since \beq
\lab{G0}\mathcal{G}(z)= \kappa\,
u^{-\frac{1}{3}}+O(u^{\frac{5}{3}})\,\,\,\,\,\,\,\,\,
\text{as}\,\,\,\,\,\, u\rightarrow0\,, \eeq
 $\mathcal{G}(z)|_{u=0}=0$ if $\kappa=0$, whereas  it diverges if $\kappa\neq0$.

For the sake of clarity, we shall analyze separately the cases
$\kappa>0$, $\kappa=0$, and $\kappa<0$.

\subsection{$\kappa>0$}\label{301}

\begin{figure}%[t]
\begin{center}
\includegraphics[width=\textwidth]{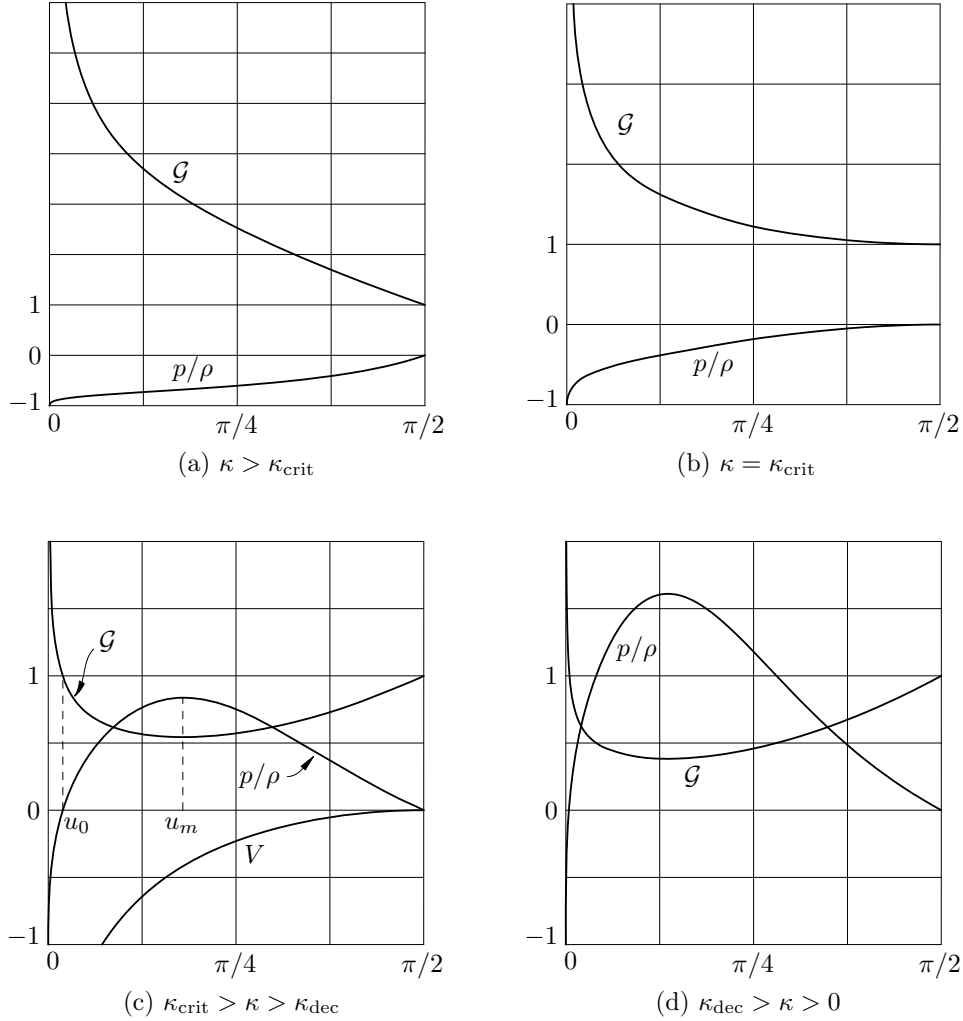}
\caption{\label{Gp}$\mathcal{G}(z)$, $V(z)$ and $p(z)$, as
functions of $u$ for decreasing values of $\kappa>0$. Since $V(z)$
is independent of $\kappa$, it is shown once.}\end{center}
\end{figure}

In this case, it is clear from \eq {G2} that  $\mathcal{G}(z)$ is
positive definite when $0<u\leq\pi/2$. On the other hand, from \eq
{gii1} and \eq {gzz1}, we get \beqa \lab {G''}
\mathcal{G}''=\mathcal{G}' V'-\mathcal{G} V''=
-\Bigl(V''+\frac{V'^2}{2}+4\pi\rho\Bigr)\mathcal{G}+4\pi C_p
= V'^2 \mathcal{G}+4\pi \rho, \eeqa%
where we have made use of \eq {gzz2}, \eq {gtt1} and  $C_p =\rho$.
Then, also $\mathcal{G}''$ is  positive definite in
$0<u\leq\pi/2$, and so $\mathcal{G}'$ is a monotonically
increasing continuous function of $u$ in this interval.

Now, taking into account that $\,\mathcal{G'}=\pd z \mathcal{G}=
\sqrt{6\pi\rho}\ \pd u\mathcal{G}$, we get from \eq {G2}  that
\beqa \mathcal{G}'(z)\lab{G'0}= -\frac{\kappa \sqrt{6\pi\rho}
}{3}u^{-\frac{4}{3}}+O(u^{\frac{2}{3}})\,\,\,\,\,\,\,\,\,
\text{as}\,\,\, u\to0, \eeqa and  from \eq {GH} that \beqa
\lab{G'1}\mathcal{G}'(z)|_{u=\pi/2}=\sqrt{6\pi\rho}\left(\kappa_{\text{crit}}-\kappa\right)\,.
 \eeqa

If $\kappa\geq \kappa_{\text{crit}}$, $\mathcal{G}'$ is negative
for small enough values of $u$ and non-positive  at ${u=\pi/2}$.
Hence $\mathcal{G}'$ is negative in $0<u<\pi/2$, so
$\mathcal{G}(z)$ is decreasing, and then
$\mathcal{G}(z)>\mathcal{G}(z)|_{u=\pi/2}=1$ in this interval (see
Fig.\ref{Gp}(a) and Fig.\ref{Gp}(b)).

For $\kappa_{\text{crit}}>\kappa>0$, $\mathcal{G}'$ is negative
for sufficiently small values of $u$ and positive at $\pi/2$. So,
there is one   (and only one) value  $u_{m}$ where it vanishes.
Clearly $\mathcal{G}(z)$ attains a local minimum  there. Hence,
there is one (and only one) value  $u_0$ ($0<u_0<\pi/2$) such that
$\mathcal{G}(z)|_{u=u_0}=\mathcal{G}(z)|_{u=\pi/2}=1$, and then
$\mathcal{G}(z)<1$ when $u_0<u<\pi/2$ (see Fig.\ref{Gp}(c)  and
Fig.\ref{Gp}(d)).

Since $\mathcal{G}(z)>0$, it is clear from \eq {p3} that $p(z)>0$
if $\mathcal{G}(z)<1$, and $p(z)$  reaches a maximum when
$\mathcal{G}(z)$ attains a minimum.

Therefore, for $\kappa\geq \kappa_{\text{crit}}$,  $p(z)$ is
negative when $0\leq u<\pi/2$ and it increases monotonically from
$-\rho$ to $0$  and it satisfies $|p|\le\rho$ all over the
space-time  (see Fig.\ref{Gp}(a) and  Fig.\ref{Gp}(b)).

On the other hand, for $\kappa_{\text{crit}}>\kappa>0$, $p(z)$
grows from $-\rho$ to a maximum positive value when $u=u_{m}$
where it starts to decrease and vanishes at $u=\pi/2$. Thus,
$p(z)$ is negative when $0<u<u_0$ and positive when $u_0<u<\pi/2$
(see Fig.\ref{Gp}(c) and Fig.\ref{Gp}(d)).  It can be readily seen
from \eq{G2} and \eq{p3} that, as $\kappa$ decreases from
$\kappa_{\text{crit}}$ to $0$, $u_m$ moves to the left and the
maximum value of $p(z)/\rho$ monotonically increases  from $0$ to
$\infty$. In section \ref{dec}, we shall show that for
$\kappa=\kappa_{dec}=0.3513\dots$ it gets $1$, and then for
$0<\kappa< \kappa_{dec}$, there is a region of space-time where
$p>\rho$ and where the dominant energy condition is thus violated.

\subsection{$\kappa=0$}

\begin{figure}[t]
\begin{center}
\includegraphics[width=\textwidth]{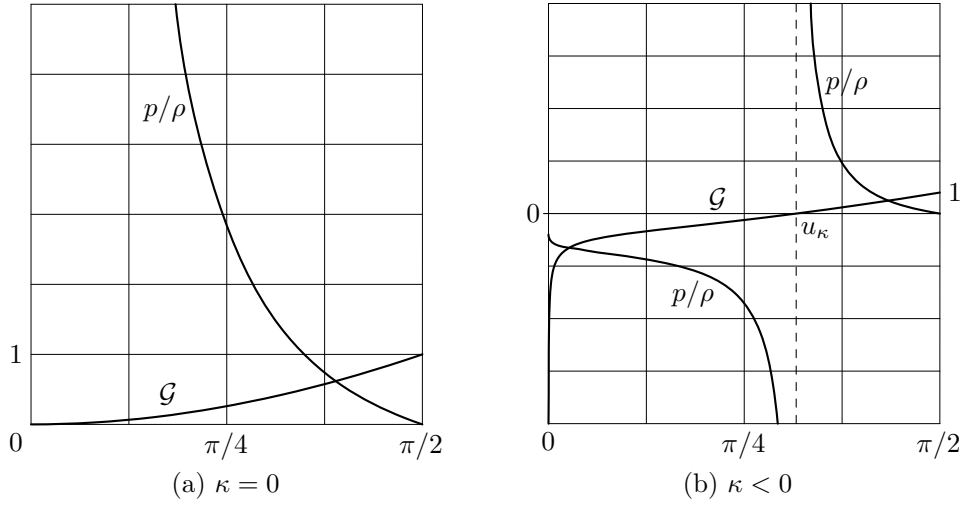}
\caption{\label{Gpneg}$\mathcal{G}(z)$ and $p(z)$ as functions of
$u$ for $\kappa\le0$.}\end{center}\end{figure}

In this case, it is clear from \eq {G2} that  $\mathcal{G}$
monotonically increases with $u$ from $0$ to
$\mathcal{G}(z)|_{u=\pi/2}=1$. Therefore, $p$ is  a monotonically
decreasing positive continuous function of $u$ in $0<u<\pi/2$ (see
Fig.\ref{Gpneg}(a)). Furthermore, at $u=0$ it diverges, since
\beqa p(z)\sim \frac{7\rho
}{3}u^{-2}\rightarrow+\infty\,\,\,\,\,\,\,\,\, \text{as}\,\,\,
u\rightarrow0. \eeqa

\subsection{$\kappa<0$}

In this case, we see from \eq{G'0} that $\mathcal{G}'$ is positive
when $u$ takes  small enough values, and from \eq{G'1} we see that
it is also positive when $u$ is near to $\pi/2$.

Now, suppose that $\mathcal{G}'(z)$ attains a local minimum when
$u=u_1$ ($0<u_1<\pi/2$), then $\mathcal{G}''(z)|_{u=u_1}=0$.
Hence, we get from \eq{G''} that  $\mathcal{G}(z)|_{u=u_1}<0$. And
taking into account that
$V'(z)|_{u=u_1}={2\sqrt{6\pi\rho}}/{3}\cot u_1>0$, we see from
\eq{gzz2}  that  $\mathcal{G}'(z)|_{u=u_1}>0$. Thus, we have shown
that $\mathcal{G}'(z)$ is a continuous positive definite function
when $0<u\leq\pi/2$ if $\kappa<0$.

Therefore, in this case, $\mathcal{G}(z)$ is a continuous function
monotonically increasing  with $u$ when $0<u\leq\pi/2$. Since it
is negative for  sufficiently small values of $u$ and  $1$ when
$u=\pi/2$ it must vanish at a unique value of $z$ when
$u=u_\kappa$ (say). Furthermore $\mathcal{G}(z)<1$ when
$0<u<\pi/2$. Clearly, we get from \eq {GH} that $u_\kappa$ is
given implicitly in terms of $\kappa$ through \beq \lab{k}{\kappa
}= \kappa_{\text{crit}}-\frac{ \,
_2F_1\!\Bigl(-\frac{1}{2},-\frac{1}{6};\frac{1}{2};\cos^2u_\kappa
\Bigr)}{{\cos u_\kappa}\;\;{\sin^{1/3} u_\kappa}} \,.\eeq We can
readily see from \eq{k} that $u_\kappa$ is a monotonically
decreasing function of $\kappa$ in $-\infty<\kappa<0$, and it
tends to $\pi/2$ when $\kappa\rightarrow-\infty$ and to $0$ when
$\kappa\rightarrow 0^{-}$.

From \eq {p3}, it is clear that $p(z)$ diverges when $u=u_\kappa$.
Furthermore, \eq {p3} also shows  that $p(z)<0$ when
$\mathcal{G}(z)<0$. And taking into account that
$\mathcal{G}(z)<1$, we see that $p(z)>0$ when $\mathcal{G}(z)>0$.
Therefore, $p(z)$ is negative when $0<u<u_\kappa$ whereas it is
positive when $u_\kappa<u<\pi/2$ (see Fig.\ref{Gpneg}(b)).

On the other hand, we see from \eq{RR} that, when $\kappa$ is
negative, another \st\, curvature singularity arises at $u_\kappa$
(besides  the one at $u=0$) since $p$ diverges there.

Therefore, if $\kappa$ is negative, the metric \eq{met2} describes
two very different space-times:

(a) For $0<u<u_\kappa$, the whole  space-time is trapped between
two singularities separated by a finite distance
$\sqrt{6\pi\rho}\,u_\kappa$. This is a space-time full of a fluid
with constant positive density $\rho$ and negative pressure $p$
monotonically decreasing with $u$,  and $p(z)|_{u=0}=-\rho$ and
$p(z)\rightarrow-\infty$ as $u\rightarrow u_\kappa$.

(b) For $u_\kappa<u<\pi/2$, the   pressure is positive and
monotonically decreasing with $u$, $p(z)\rightarrow \infty$ as
$u\rightarrow u_\kappa$ and $p(z)|_{u=\pi/2}=0$.

\section{The maximum of the pressure and the dominant energy condition}\lab{dec}

We have seen in the preceding section that for  $\kappa\geq
\kappa_{\text{crit}}$,  $p(z)$ is  negative, it increases
monotonically from $-\rho$ to $0$  and it satisfies $|p|\leq\rho$
all over the space-time  (see Fig.\ref{Gp}(a) and
Fig.\ref{Gp}(b)). On the other hand, for $\kappa\le0$, $p(z)$  is
unbounded at an inner singularity and thus the dominant energy
condition is not satisfied in this case.

For $\kappa_{\text{crit}}>\kappa>0$, since $\mathcal{G}(z)>0$, it
is clear from \eq {p}     that $p(z)>0$ if $\mathcal{G}(z)<1$, and that
$p(z)$  reaches a maximum when $\mathcal{G}(z)$ attains a minimum.
Then,  $p(z)$ grows from $-\rho$ to a maximum positive value
$p_{m}$ when $u=u_{m}$, where it starts to decrease and vanishes
at $u=\pi/2$. Thus, $-\rho\leq p(z)<0$  for $0<u\leq u_0$ and
$0<p(z)<p_m$  when $u_0<u<\pi/2$ (see Fig.\ref{Gp}(c)).

We readily see from \eq{gzz1}, since $\mathcal{G}'(z)|_{u=u_m}$
vanishes, that \beq \lab{pmax}
p_{m}=p(z)|_{u=u_m}=\frac{1}{8\pi}\left(V'(z)\right)^2|_{u=u_m}=\frac{\rho}{3}\cot^2
u_m\,,\eeq where we have made use of \eq{V}, and so the maximum
value of $p(z)$ monotonically decreases  from $\infty$ to $0$ in
$0<u_m<\pi/2$.

\begin{figure}[t]\begin{center}\includegraphics[height=4cm]{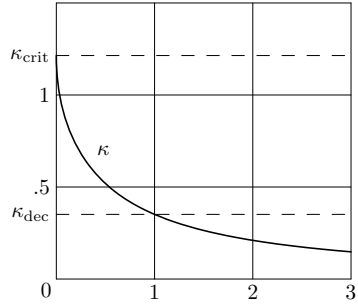}
\caption{\label{Kp}$\kappa$  as a function of
$p_m$.}\end{center}\end{figure}

Now,  by replacing \eq{pmax} into
\eq{GH} and taking into account \eq{p3}, we  can write down $\kappa$ in terms of $p_{m}$%
\beqa \lab{kappa} \kappa=\kappa_{\text{crit}}+
\frac{\left(\rho+{3p_{m}}\right)^{\frac{1}{2}}}{\left({3p_{m}}
\right)^{\frac{1}{2}}}\left(\frac{\rho^{\frac{7}{6}}}{(\rho+p_{m})(\rho+3p_{m})^{\frac{1}{6}}}-
\,_2F_1\!\left(-\frac{1}{2},-\frac{1}{6};\frac{1}{2};\frac{3p_{m}}{\rho+3p_{m}}
\right)\right)\lab{kpch} \nn\\
=\kappa_{\text{crit}}-{2}\sqrt{\frac{p_m}{3\rho}}\left(1-\frac{1}{3}\frac{p_m}{
\rho}+\frac{8 }{21 }\frac{{p_m}^3}{ \rho ^3}+\dots\right)
\text{\,\,\,for\,\,\,} p_m<\rho,\nn
\eeqa %
which clearly shows   that $\kappa\to\kappa_{\text{crit}}$  as
$p_{m}\to0$. On the other hand, by using \eq{a2}, we can write
\beq\kappa_=\frac{{2}}{\sqrt[6]{3}}\left(\frac{\rho}{p_m}\right)^{\frac{7}{6}}
\left(\frac{3}{7}-
\frac{17}{39}\left(\frac{\rho}{p_m}\right)+\dots\right)
\text{\,\,\,for\,\,\,}
p_m>\rho,\eeq%
which clearly shows  that $\kappa\to0$  as $p_{m}\to\infty$. Thus,
as $\kappa$ increases from $0$ to $\kappa_{\text{crit}}$, $p_{m}$
monotonically decreases  from $\infty$ to $0$ (see Fig.\ref{Kp}).

Hence, there is a value $\kappa_{\text{dec}}$ of $\kappa$ for
which $p_{m}=\rho$,  and from \eq{kappa} we see that it is  given
by \beq \kappa_{\text{dec}}= \kappa_{\text{crit}}+
\frac{2}{\sqrt{3}}
\left(\frac{1}{2\sqrt[3]{2}}-\;_2F_1\!\Bigl(-\frac{1}{2},-\frac{1}{6};
\frac{1}{2};\frac{3}{4} \Bigr)\right)=0.351307\dots\,.\eeq%
Also note that, in this case, we get from \eq{pmax}  that the
maximum of the pressure occurs at $u_m=\pi/6$.

Thus, for $0<\kappa< \kappa_{\text{dec}}$, there is a region of
space-time where $p>\rho$ and where the dominant energy condition
is thus violated. However, we see that for
$\kappa_{\text{dec}}\leq\kappa<\kappa_{\text{crit}}$, the
condition $|p|<\rho$ is everywhere satisfied.

Therefore, the dominant energy condition is satisfied all over the
space-time if $\kappa\geq\kappa_{\text{dec}}$.

Notice that, for $\kappa_{\text{crit}}>\kappa>0$, by eliminating
$\kappa$ by means of \eq{kappa}, the solution can be parameterized
in terms of  $p_{m}$ and $\rho$.

\section{The %$C^1$
matching of solutions and the external gravitational fields}\lab{match}

We shall discuss  matching  the interior solution to a vacuum one,
as well as  joining  two interior solutions facing each other at
the surfaces where the pressure vanishes.  For any value of
$\kappa$,  $p(z)=0$ at $u=\pi/2$, while for
$\kappa_{\text{crit}}>\kappa>0$ it also vanishes at $u=u_0$.
Therefore, the matching at $u=\pi/2$ is always possible, while the
matching at $u=u_0$ is also possible in the latter case.

We shall impose the continuity of the metric components and of
their first derivatives at the matching surfaces.

Notice that, due to the symmetry required, vacuum solutions
satisfy the field equations (\ref{gtt1}), (\ref{gii1}) and
(\ref{gzz1}), with $\rho=p=0$. In this case, we immediately get
from (\ref{gzz1}) that or $V'=0$ or $2\
\mathcal{G}'/{\mathcal{G}}+ V'=0$.

In the former case, we get from (\ref{gii1}) that $
\mathcal{G}''=0$, and the solution is \beq \lab{Rindler} ds^2=-
(A+Bz)^{2}\, dt^2 + C(dx^2+dy^2)+ dz^2\,, \eeq which is the
Rindler \st.

In the latter one, it can be written as \beq  ds^2=-
(A+Bz)^{-\frac{2}{3}}\, dt^2+C(A+Bz)^{\frac{4}{3}}\,(dx^2+dy^2)+
dz^2\,, \eeq which is the Taub's vacuum plane solution
\cite{taub}.

Therefore, as pointed out by the authors of reference \cite{NKH},
if at the matching ``plane"   the interior  $V'$ vanishes, we can
only match it to  Rindler's \st, since for the Taub's one $V'$
does not vanish at any finite point. Whereas, if on the contrary,
$V'$ does not vanish at the matching ``plane", we can only  match
the inner solution with Taub's one, since for Rindler's one, $V'$
vanishes anywhere.

Now, we see from \eq{V} that $V'$ vanishes at $u=\pi/2$ and it is
non zero at $u=u_0\neq\pi/2$. Therefore the solution can be
matched to  Ridler's \st\ at $u=\pi/2$ and to Taub's vacuum plane
solution at $u=u_0$.

Notice that in reference \cite{gs2} we did not demand the
continuity of $V'(z)$ at the matching surface and we analyzed
there the matching  of the solution to Taub's vacuum plane
solution at $u=\pi/2$.

In the next section,  we discuss the matching of the whole
interior solution to Rindler vacuum, for any value of  $\kappa$ at
$u=\pi/2$, while  we match  two interior solutions facing each
other at $u=\pi/2$ in section \ref{dos}.

In  section \ref{doble}, for $\kappa_{\text{crit}}>\kappa>0$,  we
discuss the matching of the slice  of the interior solution
$u_0\leq u\leq\pi/2$ with both vacua, while, in section
\ref{repulsive} we match the remaining piece (i.e. $0<u<u_0$) to a
Taub's vacuum.

\section{Matching the whole slab to a Rindler \st}\lab{whole}

In this section, we discuss   matching  the whole interior
solution to a vacuum one  at  $u=\pi/2$.

Since the field equations are invariant under $z$-translation, we
can choose to match the solutions at $z=0$ without losing
generality. So we select $C_2=\pi/2$, and then  \eq {GH} becomes
\beqa \lab{G3}
\mathcal{G}(z)=G_\kappa(\sqrt{6\pi\rho}\,z+\pi/2)\nn
=\frac{-\left(\kappa-\kappa_{\text{crit}}\right){\sin
(\sqrt{6\pi\rho}\,z)}+ \,
_2F_1\!\Bigl(-\frac{1}{2},-\frac{1}{6};\frac{1}{2};\sin^2(\sqrt{6\pi\rho}\,z)
\Bigr)}{{\cos^{1/3} (\sqrt{6\pi\rho}\,z)}}\,. \eeqa %
Therefore, the metric \eq{met2} reads \beqa \lab{met3} ds^2=  -
\mathcal{G}(z)^{2}\, dt^2 + \,\cos^{\frac{4}{3}}
(\sqrt{6\pi\rho}\,z)\,\left(dx^2+dy^2\right) + dz^2,\nn
-\infty<t<\infty,\quad-\infty<x<\infty,\quad-\infty<y<\infty,\quad{-\sqrt{\frac{\pi}{24\rho}}}<z\leq0\,. \eeqa%

We must impose the continuity of the components of the metric at
the matching boundary. Notice   that
$g_{tt}(0)=-\mathcal{G}(0)^2=-1$, $g_{xx}(0)=g_{yy}(0)=1$, and
$p(0)=0$.

Furthermore, we  also impose the continuity of the derivatives of
the metric components at the boundary. From \eq{G'1}, we have
\beqa \lab{G'2} \pd zg_{tt}(0)|_{\text{interior}}=
-2\,\mathcal{G}(0)\,\mathcal{G}'(0)=-2\sqrt{6\pi\rho}\left(\kappa_{\text{crit}}-\kappa\right)\,,
 \eeqa
and, from \eq{met3} we get \beqa  \lab{31} \pd
zg_{xx}(0)|_{\text{interior}}=\pd zg_{yy}(0)|_{\text{interior}}=
-\frac{4\sqrt{6\pi\rho}}{3} \cos^{\frac{1}{3}}
(\sqrt{6\pi\rho}\,z)\sin (\sqrt{6\pi\rho}\,z)\Big|_{z=0}=0\,. \nn
\eeqa

The exterior solution, i.e. for $z\geq0$, is the Rindler \st \beqa
\lab{met15} ds^2=- (1+g z)^{2}\,
 dt^2 + dx^2+dy^2+
 dz^2,\nn\nn
-\infty<t<\infty,\quad-\infty<x<\infty,\quad-\infty<y<\infty,\quad0\leq
z<\infty
\,,  \eeqa%
which  describes a homogeneous gravitational field $-g$ in the
vertical (i.e., $z$) direction.

Since $ g_{tt}(0)|_{\text{exterior}}=-1$ and $
g_{xx}(0)|_{\text{exterior}}=g_{yy}(0)|_{\text{exterior}}=1$, the
continuity of the metric components is assured. And,  concerning
the derivatives, we have \beqa \lab{gxx} \pd
zg_{xx}(z)|_{\text{exterior}}= \pd
zg_{xx}(z)|_{\text{exterior}}=0\,,\eeqa which identically matches
to \eq{31}.

Moreover, we readily get \beqa \lab{36} \pd
zg_{tt}(z)|_{\text{exterior}}= -2g \, (1+2g z)\,.\eeqa Then, by
comparing it with \eq{G'0}, we see that the continuity of $\pd
zg_{tt}$ at the  boundary yields \beqa
\lab{g}g=\sqrt{6\pi\rho}\left(\kappa_{\text{crit}}-\kappa\right)\,,\eeqa
which  relates  the  external gravitational field $g$ with matter
density $\rho$ and $\kappa$.

Now, by replacing $\kappa$ from \eq{g} into \eq{G3}, we get
 \beqa
\lab{G18} \mathcal{G}(z)=\frac{{g\,\sin
(\sqrt{6\pi\rho}\,z)}+{\sqrt{6\pi\rho}} \,\;
_2F_1\!\Bigl(-\frac{1}{2},-\frac{1}{6};\frac{1}{2};\sin^2(\sqrt{6\pi\rho}\,z)
\Bigr)}{{{\sqrt{6\pi\rho}}\;\cos^{1/3} (\sqrt{6\pi\rho}\,z)}}\,, \eeqa %
and the solution is parameterized in terms of  the external
gravitational field $g$ and the density of the matter $\rho$.

It can readily be seen from \eq{g} that, if
$\kappa>\kappa_{\text{crit}}$, $g$ is negative and  the  slab
turns out to be repulsive. If $\kappa=\kappa_{\text{crit}}$ it is
{\em gravitationally neutral}, and the exterior is one half of
Minkowski's space-time. If $\kappa<\kappa_{\text{crit}}$, it  is
attractive.

If $\kappa>0$,  the depth of the slab is
$\sqrt{\frac{\pi}{24\rho}}$ independently of the value of
$\kappa$. In this case, the pressure is  finite anywhere, but it
is negative deep below and $p=-\rho$ at the inner singularity (see
Fig.\ref{Gp}(a), Fig.\ref{Gp}(b) and Fig.\ref{Gp}(c)).  But, as
discussed in section  \ref{dec}, only when
$\kappa\geq\kappa_{dec}$ is the condition $|p|\leq\rho$ everywhere
satisfied.

If $\kappa\leq0$,  the pressure   inside the slab is always
positive, and it diverges deep below at the inner singularity (see
Fig.\ref{Gpneg}). Its depth  is \beq
d=(\pi/2-u_\kappa)/\sqrt{6\pi\rho}\,, \eeq where $u_\kappa$
($0<u_\kappa<\pi/2$) is given implicitly in terms of $\kappa$
through \eq{k}. By using \eq{k}, we can write $\kappa$ in terms of
$d$ \beqa \lab{kd} \kappa =
 \kappa_{\text{crit}}-\frac{ \,
_2F_1\!\Bigl(-\frac{1}{2},-\frac{1}{6};\frac{1}{2};\sin^2(\sqrt{6\pi\rho}\,d)
\Bigr)}{{\sin (\sqrt{6\pi\rho}\,d)}\;\;{\cos^{\frac{1}{3}}
(\sqrt{6\pi\rho}\,d)}} \,. \eeqa Now, in this case, by using
\eq{g} we can write  the external gravitational field $g$ in terms
of the matter density $\rho$ and the depth of the  slab $d$\beqa
\lab{gd} g=\frac{\sqrt{6\pi\rho}}{\sin
(\sqrt{6\pi\rho}\,d)\;\cos^{\frac{1}{3}}(\sqrt{6\pi\rho}\,d)}\,\,\,
_2F_1\!\Bigl(-\frac{1}{2},-\frac{1}{6};\frac{1}{2};\sin^2(\sqrt{6\pi\rho}\,d)
\Bigr)\,.\eeqa

For the sake of clearness, we summarize  the properties of the
solutions  discussed above in Table \ref{T1}.

\begin{table}\begin{center}
\begin{tabular}{|c|c|c|c|c|c|c|c|}   \hline
\rule{0pt}{10pt}Case&$\kappa$&$g$&  $p(z)$&$|p|\leq\rho$&Depth&Fig. \\
\hline\hline\altura

I&$\kappa>\kappa_{\text{crit}}$&$<0$&$-\rho\leq p(z)\leq0$&yes&$\sqrt{\frac{\pi}{24\rho}}$&\ref{Gp}(a)\\
\hline\altura

II&$\kappa=\kappa_{\text{crit}}$&$=0$&$-\rho\leq p(z)\leq0$&yes&$\sqrt{\frac{\pi}{24\rho}}$&\ref{Gp}(b)\\
\hline \altura

III&$\kappa_{\text{crit}}>\kappa\geq\kappa_{\text{dec}}$&$>0$&$-\rho\leq p(z)\leq p_m(\kappa)\leq\rho$&yes&$\sqrt{\frac{\pi}{24\rho}}$&\ref{Gp}(c)\\
\hline \altura

IV&$\kappa_{\text{dec}}>\kappa>0$&$>0$&$-\rho\leq p(z)\leq p_m(\kappa)$&no&$\sqrt{\frac{\pi}{24\rho}}$&\ref{Gp}(d)\\
\hline \altura

V&$0\geq\kappa$&$>0$&unbounded&no&$\frac{(\pi/2-u_\kappa)}{\sqrt{6\pi\rho}}$&\ref{Gpneg}\\
\hline

\end{tabular}\end{center}\caption{Properties of the solutions according to the value of $\kappa$.\lab{T1}}
  \end{table}

Some remarks are in order. First, notice that the maximum depth
that a slab with constant density $\rho$ can reach is
$\sqrt{\frac{\pi}{24\rho}}$, being the counterpart of the
well-known bound $M<4R/9$ ($R<\frac{1}{\sqrt{3\pi\rho}}$), which
holds for spherical symmetry.

If we  restrict ourselves to non ``exotic" matter,  the
dominant energy condition will put aside cases IV and V, as shown  in
section \ref{dec}. However, as already mentioned, it is satisfied
for the cases I, II and III (see Fig.\ref{Gp}(a), Fig.\ref{Gp}(b)
and Fig.\ref{Gp}(c)). Thus, there are still attractive, neutral
and repulsive solutions satisfying this condition.%%

 In this case, we readily get from \eq{g} the bound
 \beqa
g\leq\sqrt{6\pi\rho}\left(\kappa_{\text{crit}}-\kappa_{dec}\right)\approx3.75\,\sqrt{\rho}\,\,.\eeqa

\begin{figure}[t]
\begin{center}
\includegraphics[width=\textwidth]{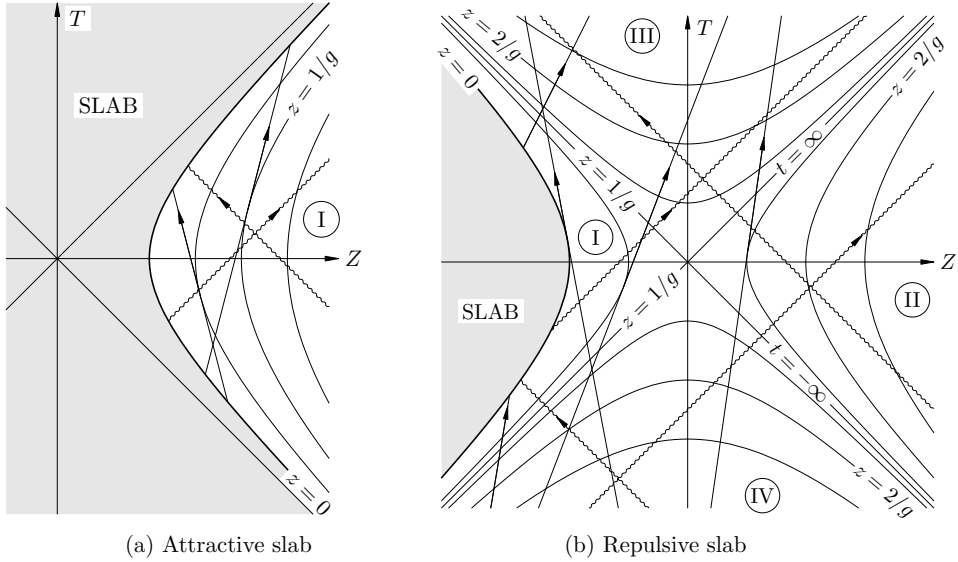}
\caption{\label{vac}Vertical time-like and null geodesics in the
vacuum }\end{center}
\end{figure}

In order to analyze the geodesics in the vacuum, it is convenient
to  consider the transformation from Rindler's coordinates $t$ and
$z$ to Minkowski's ones $T$ and $Z$ shown in Table \ref{TZ}.
Notice that, for the repulsive case, four Rindler's patches are
necessary to cover the whole exterior of the slab. Also note that,
in this case,  $z$ becomes the temporal coordinate in quadrants
III and IV, see Fig. \ref{vac}. In this coordinates, of course,
the vacuum metric becomes \beq ds^2= -dT^2+dx^2+dy^2+dZ^2\,.\eeq

Notice that, the ``planes" $z=$ constant correspond to the
hyperbolae $Z^2-T^2=$ constant, and $t=$ constant. On the other
hand, incoming vertical null geodesics are $Z+T=$ constant, and
outgoing ones are given by $Z-T=$ constant.

For attractive slabs, we readily see from  Fig. \ref{vac}(a) that
all incoming vertical photons finish at the surface of the slab,
while all outgoing ones escape to infinite. Vertical time-like
geodesics start  at the surface of the slab, reach a turning point
and fall down to the slab in a finite amount of coordinate time
$t$. Notice that a particle world-line is tangent to only one
hyperbola $Z^2-T^2= C$, with  $C>1/g$, and that the maximum value of
$z$ that it reaches is $C-1/g$.

For repulsive slabs, Fig. \ref{vac}(b), two horizons appear in the
vacuum: the lines $T=\pm Z$, showing that not all the vertical
null geodesics reach the surface of the slab. In fact, only
vertical incoming  photons coming from region IV end at the slab
surface, and only the outgoing ones finishing in region III start
at the slab surface. Incoming particles can reach the surface or
bounce at a turning point before getting it.

\begin{table}[t]
\begin{center}
\begin{tabular}{|c|c|c|c|c|c|}   \hline
\rule{0pt}{10pt}Case&Quadrant&$T$&$Z$&  $-dT^2+dZ^2$ \\
%\rule{0pt}{10pt}&&& &&&\\%everywhere&everywhere&&\\
\hline\hline

\altura Attractive&
I&$(z+{1}/{g})\sinh gt$&$(z+{1}/{g})\cosh gt$&$-(z+{1}/{g})^2dt^2+dz^2$\\
\hline\hline

\altura&I and II&$(z-{1}/{g})\sinh gt$&$(z-{1}/{g})\cosh gt$&$-(z-{1}/{g})^2dt^2+dz^2$\\
\cline{2-5}

\altura Repulsive&III&$(z-{1}/{g})\cosh gt$&$(z-{1}/{g})\sinh gt$&$-dz^2+(z-{1}/{g})^2dt^2$\\
\cline{2-5}

\altura&IV&$({1}/{g}-z)\cosh gt$&$(z-{1}/{g})\sinh gt$&$-dz^2+(z-{1}/{g})^2dt^2$\\
\hline
\end{tabular}\end{center}\caption{The  transformations from Rindler's coordinates to Minkowski's ones. \lab{TZ} }
  \end{table}

\section{Matching two slabs}\lab{dos}

Now we consider two incompressible fluids joined at $z=0$ where
the pressure vanishes, the lower one having  density $\rho$ and
the upper having  density $\rho'$. Thus, the lower solution is
given by \eq{met3}. By means of the transformation
$z\rightarrow-z$, $\rho\rightarrow\rho'$ and
$\kappa\rightarrow\kappa'$ we  get the upper one \beqa
\lab{met4} ds^2=  -
G_{\kappa'}(\pi/2-\sqrt{6\pi\rho'}\,z)^{2}\, dt^2 +
\,\cos^{\frac{4}{3}} (\sqrt{6\pi\rho'}\,z)\,\left(dx^2+dy^2\right)
+ dz^2,\nn
-\infty<t<\infty,\quad-\infty<x<\infty,\quad-\infty<y<\infty,\quad 0\leq z<\sqrt{\frac{\pi}{24\rho'}}\,. \eeqa%
From \eq{met3}, \eq{31}  and \eq{met4}, we can readily see that
$g_{tt}(z)$, $g_{xx}(z)$ and $\pd zg_{xx}(z)$ are continuous at
$z=0$. Furthermore, from \eq{G'2} we see that the continuity of
$\pd zg_{tt}$ requires \beqa
\lab{G'4}\sqrt{\rho}\left(\kappa_{\text{crit}}-\kappa\right)=-\sqrt{\rho'}
\left(\kappa_{\text{crit}}-\kappa'\right)\,.
 \eeqa
Thus, if one solution has a $\kappa$ greater than
$\kappa_{\text{crit}}$, the other one must have it smaller than
$\kappa_{\text{crit}}$. Therefore, the joining is only possible
between an attractive solution and a repulsive one, or between two
neutral ones.

It is easy to see that we can also insert  a slice of arbitrary
thickness of the vacuum solution \eq{taub} between them, obtaining
a full relativistic plane ``gravitational capacitor". For example,
we can trap a slice of Minkowski's \st\, between two solutions
with $\kappa=\kappa_{\text{crit}}$.

\section{Attractive Slab surrounded by two different vacuums } \lab{doble}

We have already seen that, in the case
$\kappa_{\text{crit}}>\kappa>0$, the pressure  also vanishes
inside the slab at the point where $u=u_0$. Here we discuss the
matching of  the slice  of the interior solution $u_0\leq
u\leq\pi/2$ with two vacuums.

Clearly, the thickness of the slab $d$ is given by
 \beq
\lab{d}d=\frac{(\pi/2-u_0)}{\sqrt{6\pi\rho}}\,,\eeq and
$0<d<\sqrt{\frac{\pi}{24\rho}}$.

\begin{figure}[t]\begin{center}\includegraphics[height=5.2cm]{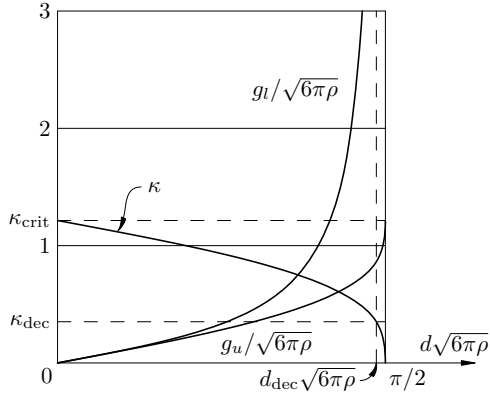}
\caption{\label{kgg}$\kappa$, $g_u$ and $g_l$  as  functions of
$d$.}\end{center}\end{figure}

Since $\mathcal{G}(z)|_{u=u_0}=1$, we can write down from \eq{GH}
the  expression which gives  $\kappa$ in terms of $d$  and $\rho$
\beqa \lab{kdd} \kappa=\kappa_{\text{crit}}+
\frac{{\cos^{1/3}(\sqrt{6\pi\rho}\,d)}- \,
_2F_1\!\Bigl(-\frac{1}{2},-\frac{1}{6};\frac{1}{2};\sin^2(\sqrt{6\pi\rho}\,d)
\Bigr)}{{\sin (\sqrt{6\pi\rho}\,d)}}\\ =
\kappa_{\text{crit}}-\frac{\sqrt{6\pi\rho}\,d}{3}\left(1+\frac{2\pi\rho}{3}d^2+\frac{(2\pi\rho)^2}{5}d^4+\dots\right)
\text{\,\,\,for\,\,\,} \sqrt{6\pi\rho}\,d<1\,, \lab{kdch} \eeqa
which clearly shows   that $\kappa\to\kappa_{\text{crit}}$  as
$d\to0$. On the other hand, by using \eq{a2}, we can write
\beq\kappa_=\frac{{\cos^{1/3}(\sqrt{6\pi\rho}\,d)} }{{\sin
(\sqrt{6\pi\rho}\,d)}}
\left(1-\frac{3}{7}\cos^{2}(\sqrt{6\pi\rho}\,d)+\dots\right)
%\text{\,\,\,for\,\,\,}\frac{\pi}{2}-\sqrt{6\pi\rho}\,d<1,
\eeq%
which clearly shows  that $\kappa\to0$  as
$d\to\sqrt{\frac{\pi}{24\rho}}$. Thus, as $d$ increases from $0$
to $\sqrt{\frac{\pi}{24\rho}}$,   $\kappa$ monotonically decreases
from $\kappa_{\text{crit}}$ to $0$ (see Fig.\ref{kgg}).

Therefore, the maximum thickness $d_{\text{dec}}$ that a solution
satisfying  the dominant energy condition  can have, satisfies
\beqa \lab{kddec} \kappa_{\text{dec}}=\kappa_{\text{crit}}+
\frac{{\cos^{1/3}(\sqrt{6\pi\rho}\,d_{\text{dec}})}- \,
_2F_1\!\Bigl(-\frac{1}{2},-\frac{1}{6};\frac{1}{2};\sin^2(\sqrt{6\pi\rho}\,d_{\text{dec}})
\Bigr)}{{\sin (\sqrt{6\pi\rho}\,d_{\text{dec}})}} \,. \eeqa A
straightforward numerical computation gives
$\sqrt{6\pi\rho}\,d_{\text{dec}}=1.52744\dots$. Therefore, if
$0<d<d_{\text{dec}}$, the dominant energy condition is satisfied
anywhere. Whereas if $d_{\text{dec}}<d<\sqrt{\frac{\pi}{24\rho}}$,
there is a region inside the slab where $p(z)>\rho$.

Now, by eliminating $\kappa$ by means of \eq{kdd}, the solution
can be parameterized in terms of  $d$ and $\rho$, and \eq{G3}
becomes \beqa \lab{G4} \mathcal{G}(z)=\frac{
\,_2F_1\!\Bigl(-\frac{1}{2},-\frac{1}{6};\frac{1}{2};\sin^2(\sqrt{6\pi\rho}\,z)
\Bigr)}{{\cos^{1/3} (\sqrt{6\pi\rho}\,z)}}
-\frac{{\cos^{\frac{1}{3}} (\sqrt{6\pi\rho}\,d)}}{\sin
(\sqrt{6\pi\rho}\,d)} \, \frac{\sin
(\sqrt{6\pi\rho}\,z)}{\cos^{\frac{1}{3}} (\sqrt{6\pi\rho}\,z )}
\nn +\ \frac{{
_2F_1\!\Bigl(-\frac{1}{2},-\frac{1}{6};\frac{1}{2};\sin^2(\sqrt{6\pi\rho}\,d)\Bigr)}}{\sin
(\sqrt{6\pi\rho}\,d)}\, \frac{\sin
(\sqrt{6\pi\rho}\,z)}{\cos^{\frac{1}{3}}
(\sqrt{6\pi\rho}\,z )} \, . \eeqa %
Notice that it clearly  shows that
$\mathcal{G}(-d)=\mathcal{G}(0)=1$. By means of  \eq{p3} and
\eq{G4}  $p(z)$ can also be explicitly  written down in
terms  of  $d$ and $\rho$. The inner line element \eq{met3} reads
\beqa \lab{met5} ds^2= - \mathcal{G}(z)^{2}\, dt^2 +
\,\cos^{\frac{4}{3}} (\sqrt{6\pi\rho}\,z)\,(dx^2+dy^2) +
dz^2,\nn
-\infty<t<\infty,\quad-\infty<x<\infty,\quad-\infty<y<\infty,\quad{-\sqrt{\frac{\pi}{24\rho}}}<-d\leq z\leq0\,. \eeqa%

We must impose the continuity of the components of the metric and
their first derivatives at both matching boundaries, i.e. $z=0$
and $z=-d$.

The matching at $z=0$ was already discussed in section
\ref{doble}. Thus, the upper exterior solution, i.e. for $z\geq0$,
is the Rindler's \st \beqa \lab{met6} ds^2=- (1+g_uz)^{2}\,
 dt^2 + dx^2+dy^2+
 dz^2,\nn\nn
-\infty<t<\infty,\quad-\infty<x<\infty,\quad-\infty<y<\infty,\quad0\leq
z<\infty
\,,  \eeqa%
which   describes a homogeneous gravitational field $-g_u$ in the
vertical (i.e., $z$) direction. And, according to \eq{g}, we see
that the continuity of $\pd zg_{tt}$ at the upper boundary yields
\beqa
\lab{gu}g_u=\sqrt{6\pi\rho}\left(\kappa_{\text{crit}}-\kappa\right)\,,\eeqa
which  relates  the upper external gravitational field $g_u$ with
matter density $\rho$ and $\kappa$. By using \eq{kdd}, we can also
write it in terms of $d$ and $\rho$ \beqa \lab{gud}
g_u=\frac{\sqrt{6\pi\rho}}{{\sin (\sqrt{6\pi\rho}\,d)}}\left( { \,
_2F_1\!\Bigl(-\frac{1}{2},-\frac{1}{6};\frac{1}{2};\sin^2(\sqrt{6\pi\rho}\,d)
\Bigr)}-{\cos^{\frac{1}{3}}(\sqrt{6\pi\rho}\,d)}\right)\,.\eeqa

At the lower boundary, we have $g_{tt}(-d)=-\mathcal{G}(-d)^2=-1$,
$g_{xx}(-d)=g_{yy}(-d)=\cos^{\frac{4}{3}} (\sqrt{6\pi\rho}\,d)$,
and  $p(-d)=0$.

 On the other hand, regarding the derivatives, since $\mathcal{G}(z)|_{u=u_0}=1$ and
$p(z)|_{u=u_0}=0$, from \eq{gzz1} we get
 \beqa \lab{G'22} \mathcal{G}'(z)|_{u=u_0}=-\frac{1}{2}V'(z)|_{u=u_m}=
  -\frac{\sqrt{6\pi\rho}}{3}\,\cot u_0= -\frac{\sqrt{6\pi\rho}}{3}\,\tan(\sqrt{6\pi\rho}\,d)\,,
 \eeqa
where we have made use of \eq{V} and \eq{d}. Thus,  \beqa
\lab{G'd} \pd zg_{tt}(-d)|_{\text{interior}}=
-2\,\mathcal{G}(-d)\,\mathcal{G}'(-d)=2\frac{\sqrt{6\pi\rho}}{3}\,\tan(\sqrt{6\pi\rho}\,d)\,.
 \eeqa
While from \eq{met5}, we get
  \beqa \lab{gxx'd}  \pd zg_{xx}(-d)|_{\text{interior}}=\pd zg_{yy}(-d)|_{\text{interior}}
=4\frac{\sqrt{6\pi\rho}}{3}\,\cos^{\frac{1}{3}}
(\sqrt{6\pi\rho}\,z)\,\sin(\sqrt{6\pi\rho}\,d)\,.
 \eeqa

Taking into account the discussion in section \ref{match}, we can
write the corresponding lower exterior solution, i.e. for $z<-d$,
as \beqa \lab{met7}  ds^2=-
\left(1+3g_l(d+z)\right)^{-\frac{2}{3}}\,
 dt^2 + C_d\,\left(1+3g_l(d+z)\right)^{\frac{4}{3}}\,(dx^2+dy^2) +
 dz^2,\nn
-\infty<t<\infty,\quad-\infty<x<\infty,\quad-\infty<y<\infty,\quad-d-\frac{1}{3g_l}<
z\leq -d
\,,  \eeqa%
which  describes a homogeneous gravitational field $+g_l$ in the
vertical  direction and finishes up at an empty singular boundary
at $z=-d-\frac{1}{3g_l}$.

Since $ g_{tt}(-d)|_{\text{exterior}}=-1$ and $
g_{xx}(-d)|_{\text{exterior}}=g_{yy}(-d)|_{\text{exterior}}=C_d$,
we see that, taking into account \eq{met3}, the continuity of the
metric components is assured if we set $C_d=\cos^{\frac{4}{3}}
(\sqrt{6\pi\rho}\,d)$.  And, concerning the derivatives of
metric's components, we have\beqa \lab{G'3} \pd
zg_{tt}(z)|_{\text{exterior}}= 2g_l\,
\left(1+3g_l(d+z)\right)^{-\frac{5}{3}}\,.
 \eeqa
Therefore,   by comparing with \eq{G'd}, we see that the
continuity of  $\pd zg_{tt}$ at the lower boundary yields \beqa
\lab{gl}g_l=\frac{\sqrt{6\pi\rho}}{3}\,\tan(\sqrt{6\pi\rho}\,d)\,,\eeqa
which  relates  the lower external gravitational field $g_l$ with
$d$ and $\rho$.

On the other hand, we get from \eq{met7} \beqa \lab{gxx'd1} \pd
zg_{xx}(z)|_{\text{exterior}}=\pd zg_{yy}(z)|_{\text{exterior}}
=4\,\cos^{\frac{4}{3}} (\sqrt{6\pi\rho}\,d)\,g_l\,
\left(1+3g_l(d+z)\right)^{\frac{1}{3}}\,. \eeqa Taking into
account \eq{gl}, by comparing the last equation with \eq{gxx'd},
we see that the matching is complete.

This solution is remarkably asymmetric, not only because both
external gravitational fields are different, as we can readily see
by comparing \eq{gud} and \eq{gl} (see also Fig.\ref{kgg}), but
also because the nature of vacuums is completely different: the
upper one is flat and semi-infinite, whereas the lower one is
curved and finishes up down bellow at an empty repelling boundary
where space-time curvature diverges.

This exact solutions clearly show how the attraction of distant
matter can shrink the \st\ in such a way that it finishes at an
empty singular boundary, as pointed out in \cite{gs}.

Free particles or photons move in the lower vacuum
($-d-\frac{1}{3g_l}<z<-d$) along the time-like or null geodesics
discussed in detail in \cite{gs}. So,  all geodesics start  and
finish at the boundary of the slab and have a turning point.
Non-vertical geodesics  reach a turning point point at a finite
distance from the   singularity (at $z=-d-\frac{1}{3g_l}$), and
the smaller their horizontal momentum is, the closer they get the
singularity. The same occurs for vertically moving particles, i.
e., the higher the energy, the closer they approach the
singularity. Only vertical null geodesics just touch the
singularity and bounce (see Fig. 1 of Ref. \cite{gs} upside down).

\section{The Newtonian limit and the  restoration of the mirror
symmetry}\lab{asymmetry}

It should be noted that the solutions so far discussed are
mirror-asymmetric. In fact, it has been  shown in \cite{NKH} that
the solution cannot have a ``plane" of symmetry in a region where
$p(z)\geq0$. In order to see this, suppose that $z=z_s$ is that ``plane", then it
must  hold that $ \mathcal{G}'= V' =p' =0$ at $z_s$, and so from
\eq {gzz1} we get that also $p(z_s)=0$, and then
$\mathcal{G}(z_s)=1$. Now, by differentiating \eq{ppr} and using
\eq{G''}, we obtain $p''(z_s)=-4\pi\rho^2<0$.

Notice that \eq{V} and the condition $V'(z_s)=0$ imply that
$u=\pi/2$. And  then we get from \eq{G'1} that the condition
$\mathcal{G}'(z_s)=0$ implies $\kappa=\kappa_{\text{crit}}$.

Therefore, the only mirror symmetric solutions is the joining of
two identical neutral slabs discussed in section \ref{dos}. We
clearly get from \eq{G3}, that for this solution we have \beqa
\lab{Gsymm} \mathcal{G}(z) = \frac{ \,
_2F_1\!\Bigl(-\frac{1}{2},-\frac{1}{6};\frac{1}{2};\sin^2(\sqrt{6\pi\rho}\,z)
\Bigr)}{{\left(1-\sin^2(\sqrt{6\pi\rho}\,z)\right)^{1/6}  }}\,,
\eeqa which shows that it is a $C^\infty$ even function of $z$ in
$-\sqrt{\frac{\pi}{24\rho}}<z<\sqrt{\frac{\pi}{24\rho}}$. But, of
course,  we have seen in section \ref{solution} that $-\rho\leq
p(z)\leq0$ in this case.

However, for the solution of the preceding section, this asymmetry
turns out to disappear when $\sqrt{6\pi\rho}\,d\ll1$.  In fact,
from \eq{gud} we get \beqa \lab{gud1}g_u=2\pi\rho\, d ( 1+
\frac{2}{3}\pi\rho\, d^2 + \dots)\text{\,\,\,for\,\,\,}
\sqrt{6\pi\rho}\,d<1\,,\eeqa while from \eq{gl} we get \beqa
\lab{gl1}g_l=2\pi\rho\, d ( 1+ 2\pi\rho\, d^2 +
\dots)\text{\,\,\,for\,\,\,} \sqrt{6\pi\rho}\,d<1\,.\eeqa Hence,
both gravitational fields tend to the Newtonian result $2\pi\rho\,
d$, and the difference between them is of the order
$(\sqrt{6\pi\rho} \,d)^3$.

Furthermore, in this limit, \eq{G4} becomes \beqa \lab{G5}
\mathcal{G}(z)= 1 + 2\pi\rho z(z+d)+ \frac{4}{3} \pi ^2 \rho ^2
z(z^3+d^3)+O((\sqrt{6\pi\rho}\,d)^6)
\, , \eeqa %
so
\beqa  g_{tt}(z)=- \mathcal{G}(z)^2\approx -\left(1 + 4\pi\rho z(z+d)\right)\, , \eeqa %
which shows that the Newtonian potential inside the slab tends to
\beqa  \Phi(z)=2\pi\rho z(z+d)\, . \eeqa %
Since $\Phi(-\frac{d}{2}-z)=\Phi(-\frac{d}{2}+z)$, it is
mirror-symmetric at $z=-d/2$.

Moreover, we obtain from \eq{p} that  the pressure inside the slab
tends to the hydrostatic Newtonian result
\beqa  p(z)=-2\pi\rho^2 z(z+d)\, . \eeqa %
It should also be noted, by comparing \eq{kpch} and \eq{kdch},
that in this limit, they lead to
 \beq {\frac{p_m}{\rho}}=\frac{\pi\rho}{2}\,d^2\,.\eeq

Therefore, in the Newtonian limit, the the mirror symmetry at the
middle point of the slab is restored.

\section{Thinner Repelling Slabs }\lab{repulsive}

By exchanging the place of  matter and vacuum, we can also match
the piece of the interior solution discarded in section
\ref{doble}  to the discarded  asymptotically flat tail of  Taub's
vacuum,  thus getting a repulsive slab.

Clearly, the inner solution is given by \eq{met5}, but now
$-\sqrt{\frac{\pi}{24\rho}}<z\leq-d$. While the outer one is given
by \eq{met7} with $-d\leq z$. Therefore, %since the solution is the
%same as the one of section \ref{doble},
we get from \eq{gl} \beqa
g=\frac{\sqrt{6\pi\rho}}{3}\,\tan(\sqrt{6\pi\rho}\,d)\,,\eeqa
which  relates  the  external gravitational field $g$ with $d$ and
$\rho$. But now, the thickness of this slab is \beq \lab{d'}
d^\prime=
\sqrt{\frac{\pi}{24\rho}}-d<\sqrt{\frac{\pi}{24\rho}}\,,\eeq and
so \beqa
g=\frac{\sqrt{6\pi\rho}}{3}\,\cot(\sqrt{6\pi\rho}\,d')\,.\eeqa

Of course, by means of  \eq{p3}, \eq{G4} and \eq{d'},
$\mathcal{G}(z)$ and $p(z)$ can also be explicitly  written down
in terms  of  $d'$ and $\rho$ in this case.

In this repulsive case, free particles or photons move in the
vacuum ($z>0$)  along the  mirror image of time-like or null
geodesics discussed in detail in \cite{gs}. All occurs in the
vacuum as if there were  a Taub singularity inside the matter at a
distance  $|1/3g|$ from the surface---this image singularity
should not be confused with the ``real" inner one situated $d'$
from the surface. Therefore, only  the Taub's geodesics  for which
the distance between the turning point and the image singularity
is smaller than $\frac{1}{3g}$, should be cut at slab's surface.
For instance, this always occurs for vertical photons. These facts
are easily seen  by looking at Fig. 1 of reference \cite{gs}
upside down and by exchanging the position of vacuum and matter.

Notice that these slabs turn out to be less repulsive than the
ones discussed in section \ref{whole}, since all incoming vertical
null geodesics reach the slab surface in this case.

\section{Concluding remarks }

 We have done a detailed study of  the exact solution of
Einstein's equations corresponding to a static and plane symmetric
distribution of  matter with constant  positive density. By
matching this internal solution to  vacuum ones,  we showed that
different situations arise depending on the value of a parameter
$\kappa$.

We found that  the dominant energy condition is satisfied only for
$\kappa\geq\kappa_{dec}=0.3513\dots$.

As a result of the matching, we get very simple complete (matter
and vacuum) exact solutions presenting some somehow astonishing
properties without counterpart in Newtonian gravitation:

The maximum depth that these slabs can reach is
$\sqrt{\frac{\pi}{24\rho}}$ and the solutions turn out to be
remarkably asymmetric.

We found repulsive slabs in which negative but bounded ($|p|\leq
\rho$) pressure dominate the attraction of the matter. These
solutions finish deep below at a singularity where $p=-\rho$. If
their depth is smaller than $\sqrt{\frac{\pi}{24\rho}}$, the
exterior is the asymptotically flat tail of Taub's vacuum plane
solution, while when they reach the maximum depth the vacuum turns
out to be a flat Rindler space-time with event horizons, showing
that there are incoming vertical photons which never reach the
surface of the slabs in this case.

We also found attractive solutions finishing deep below at a
singularity. In this case the outer solution in this case is a
Rindler \st.

We also described a non-singular solution of thickness $d$
surrounded by two vacuums. This solution turns out to be
attractive and remarkably asymmetric because the nature of both
vacuums is completely different: the ``upper" one is flat and
semi-infinite, whereas the ``lower" one is curved and finishes up
down below at an empty repelling boundary where space-time
curvature diverges. The pressure is positive and bounded,
presenting a maximum at an asymmetrical position between  the
boundaries. We explicitly wrote down the pressure  and the
external gravitational fields in terms of $\rho$ and  $d$. We show
that if $0<\sqrt{6\pi\rho}\,d<1.52744\dots$, the dominant energy
condition is satisfied all over the space-time. We also  show how
the mirror symmetry is restored at the Newtonian limit. These
exact solutions clearly show how the attraction of distant matter
can shrink the \st\ in such a way that it finishes at an empty
singular boundary, as pointed out in \cite{gs}.

We have also discussed  matching an attractive slab to a repulsive
one, and two neutral ones. We also comment on how to assemble
relativistic gravitational capacitors consisting of a slice of
vacuum trapped between two  such slabs.

\section*{Appendix: Some properties of $_2F_1(a,b;c;x)$}

Here, we show  how  the integral appearing in the first line of
\eq{integral} is performed. By doing the change of variable
$t=\sin^2 u'$, we can write \beq  \int_0^u \sin^a u'\,\cos^b
u'\, du' = \frac{1}{2}\int_0^{\sin^2 u }t^{\frac{a-1}{2}}\,
(1-t)^{\frac{b-1}{2}}\, dt=\frac{1}{2}\ B_{\sin^2 u
}\left(\frac{a+1}{2},\frac{b+1}{2}\right)\,,\eeq where $B_x(p,q)$
is the incomplete beta function, which is related to a
hypergeometric function through \beq B_x(p,q)=\frac{x^p}{p}\,
_2F_1(p,1-q;p+1;x) \eeq (see for example \cite{tablarusa}).
Therefore, \beqa \int_0^u \sin^a u'\,\cos^b u'\, du' = \frac{(\sin
u)^{a+1}}{a+1}\
_2F_1\!\Bigl(\frac{a+1}{2},\frac{1-b}{2};\frac{a+3}{2};\sin^2u\Bigr)\nn
=\frac{1}{a+1}\ (\sin u)^{a+1} \ (\cos
u)^{b+1}\:\:_2F_1\!\Bigl(1,\frac{a+b+2}{2};\frac{a+3}{2};\sin^2u\Bigr)
\,,\eeqa where we used the transformation
$_2F_1(a,b;c;x)=(1-x)^{c-a-b}\, {}_2F_1(c-a,c-b;c;x)$ in the last
step.

For the sake of completeness,  we  display here the very few
formulas involving   hypergeometric functions $_2F_1(a,b;c;z)$
required to  follow through all the steps of this paper.

As it is well known \beqa
_2F_1(a,b;c;z)=1+\frac{ab}{c}z+\frac{a(a+1)b(b+1)}{c(c+1)}\frac{z^2}{2!}+\dots\,,\text{\,\,\,for\,\,\,}
|z|<1\,.\eeqa

By using the transformation \cite{tablarusa,abra}
\beqa_2F_1(a,b;c;z)=\frac{\Gamma(c)\Gamma(c-a-b)}{\Gamma(c-a)\Gamma(c-b)}\;\;
{}_2F_1(a,b;a+b-c+1;1-z)\nn
+\,(1-z)^{c-a-b}\;\frac{\Gamma(c)\Gamma(a+b-c)}{\Gamma(a)\Gamma(b)}\;\;
{}_2F_1(c-a,c-b;c-a-b+1;1-z)\,,\eeqa with $a=-1/2$, $b=-1/6$, and
$c=1/2$, we find the useful relations \beqa\lab{a1}
\frac{3}{7}\,\,(1-z)^{\frac{7  }{6 }}\,\,
_2F_1\!\Bigl(1,\frac{2}{3};\frac{13}{6};1-z \Bigr)=
-\frac{\sqrt{\pi}\,{\Gamma(\frac{7}{6})}}{{\Gamma(\frac{2}{3})}}\,
\sqrt{z}\,+\; _2F_1\!\Bigl(-\frac{1}{2},-\frac{1}{6};\frac{1}{2};z
\Bigr)\\= -\frac{ \sqrt{\pi } \Gamma \left(\frac{7}{6}\right) }{
\Gamma
   \left(\frac{2}{3}\right)}\,\sqrt{z}\,+1+\frac{z}{6}+\frac{5 z^2}{216}+\frac{11 z^3}{1296}+\dots\,,\text{\,\,\,for\,\,\,}
|z|<1\,, \eeqa or, by making $z\to 1-z$, \beqa  \;
_2F_1\!\Bigl(-\frac{1}{2},-\frac{1}{6};\frac{1}{2};1-z \Bigr)
=\frac{\sqrt{\pi}\,{\Gamma(\frac{7}{6})}}{{\Gamma(\frac{2}{3})}}
\sqrt{1-z}+\frac{3}{7}\,z^{\frac{7}{6}}\,
_2F_1\!\Bigl(1,\frac{2}{3};\frac{13}{6};z \Bigr)\\ =\frac{
\sqrt{\pi } \Gamma \left(\frac{7}{6}\right) }{ \Gamma
   \left(\frac{2}{3}\right)}\,\sqrt{1-z}+\frac{3  }{7 }\,
z^{\frac{7  }{6 }}\left(1+\frac{4z}{13}+\frac{40
z^2}{247}+\frac{128 z^3}{1235}+\dots
\right)\,,\text{\,\,\,for\,\,\,} |z|<1\,.\lab{a2} \eeqa

%\section*{References}
%%%%%%%%%%%%%

\end{document}